\documentclass{esagnc}

\graphicspath{{./images/}}

\title{Can Julia land on the Moon? \\On the development of a GNC simulation \\framework for the Argonaut lunar lander}

\author[1,*]{Francesco Capolupo}
\author[1]{Frederik Markus}

\affil[1]{European Space Agency, Keplerlaan 1, Noordwijk 2201 AZ, The Netherlands}
\affil[*]{Corresponding author: francesco.capolupo@esa.int}

\authorsInShort{F. Capolupo and F. Markus}
\copyrgt{European Space Agency}

\usepackage{subcaption}
\usepackage{xcolor}
\usepackage[most]{tcolorbox}
\usepackage{wrapfig}
\usepackage{booktabs}

\usepackage[T1]{fontenc}
\usepackage{inconsolata}
\usepackage{listings}
\tcbuselibrary{listings}

\definecolor{codebg}{rgb}{0.16,0.18,0.21}
\definecolor{dracFg}{HTML}{F8F8F2}
\definecolor{dracComment}{HTML}{6272A4}
\definecolor{dracCyan}{HTML}{8BE9FD}
\definecolor{dracGreen}{HTML}{50FA7B}
\definecolor{dracOrange}{HTML}{FFB86C}
\definecolor{dracPink}{HTML}{FF79C6}
\definecolor{dracPurple}{HTML}{BD93F9}
\definecolor{dracRed}{HTML}{FF5555}
\definecolor{dracYellow}{HTML}{F1FA8C}

\lstdefinelanguage{Julia}{
  morekeywords=[1]{
    function,end,struct,mutable,return,if,else,elseif,for,while,
    do,try,catch,finally,break,continue,const,global,local,
    import,using,export,module,baremodule,quote,let,begin,
    abstract,type,primitive,where,in,isa,typeof
  },
  morekeywords=[2]{
    Float64,Int64,Int,Bool,String,Any,Vector,Union,@inline,@inbounds,
  },
  sensitive=true,
  morecomment=[l]{\#},
  morestring=[b]",
  alsoletter={_,!,?,@},
  literate=
    {::}{{\textcolor{dracPink}{::}}}2
    {=}{{\textcolor{dracPink}{=}}}1
    {<}{{\textcolor{dracPink}{<}}}1
    {>}{{\textcolor{dracPink}{>}}}1,
}

\lstdefinestyle{mycode}{
  language=Julia,
  backgroundcolor=\color{codebg},
  basicstyle=\footnotesize\ttfamily\color{dracFg},
  keywordstyle=[1]\color{dracPink}\bfseries,
  keywordstyle=[2]\color{dracCyan},
  keywordstyle=[3]\color{dracPink},
  commentstyle=\color{dracComment}\itshape,
  stringstyle=\color{dracYellow},
  identifierstyle=\color{dracFg},
  numberstyle=\color{dracPurple},
  breaklines=true,
  breakatwhitespace=false,
  tabsize=4,
  showstringspaces=false,
  columns=fullflexible,
  keepspaces=true,
  frame=none,
}

\tcbset{
  mycode/.style={
    enhanced,
    boxrule=0pt,
    colback=codebg,
    arc=8pt,
    listing only,
    listing options={style=mycode},
  }
}

\begin{document}

\maketitle

\begin{abstract}
    No, the Julia programming language cannot land on the Moon — but it can play a crucial role in designing and analysing the Guidance, Navigation, and Control (GNC) algorithms required for doing so. This paper presents the development of a lunar landing simulation framework implemented in Julia at the European Space Agency (ESA), within the Argonaut lunar lander programme.
    ATLAS (Argonaut Tools for Landing Analysis and Simulation) is a modular suite of analysis and simulation tools that cover the complete descent and landing phase of Argonaut, integrating high fidelity translational and rotational dynamics, varying mass properties, propellant sloshing, detailed sensor and actuator models, and flight-representative GNC algorithms within a multi-rate simulation environment. The framework is intended to bridge early-phase prototyping and large-scale Monte Carlo analysis within a single environment. This work evaluates the advantages and limitations of adopting Julia compared to established GNC development practices based on the MATLAB/Simulink ecosystem. The results show that Julia provides a powerful, flexible, and high-performance environment for agency-driven research, early-phase design studies, and computationally intensive closed-loop simulations enabling large-scale, parallelizable simulations and rapid design iteration cycles.
\end{abstract}

\section{Introduction}

The Argonaut Lander is a key element of ESA's Terrae Novae Exploration Programme. Scheduled to begin operations from 2030 onwards, its Lunar Descent Element (LDE) is a $\sim$10-ton lander capable of delivering up to 1500 kg of scientific or logistic payload to the lunar surface with pin-point landing accuracy. This capability will support a broad range of missions, including cargo delivery for the Artemis programme, sample return, and scientific and technology demonstration missions.

The development of the lander requires extensive Guidance, Navigation and Control (GNC) analysis throughout all project phases, relying heavily on high-fidelity closed-loop simulations to assess design maturity, evaluate performance, characterize uncertainties, and support key engineering decisions. In addition to its internal assessment activities, ESA requires independent simulation capabilities to provide technical support to the industrial prime contractor during the design, development, and verification of the GNC subsystem. Such simulators play a central role in consolidating architectural choices, assessing system performance, and conducting large-scale sensitivity and Monte Carlo analyses.

Traditionally, ESA and its industrial partners have relied on simulation frameworks developed in MATLAB and Simulink for these activities. These tools have become the de facto standard within the space industry thanks to their extensive numerical libraries, mature graphical modelling environment, and widespread user familiarity. However, the growing need for large Monte Carlo campaigns and rapid design iterations, exposes some of the limitations of these traditional environments, particularly in terms of simulation speed and ability to easily interface with other software, such as image renderers. Rather than developing yet another simulator on MATLAB/Simulink, it was decided to use Argonaut as an opportunity to explore alternative approaches and investigate whether the use of a different programming language could provide tangible benefits.

\begin{wrapfigure}{r}{0.4\textwidth} %this figure will be at the right
    \centering
    \includegraphics[width=0.4\textwidth]{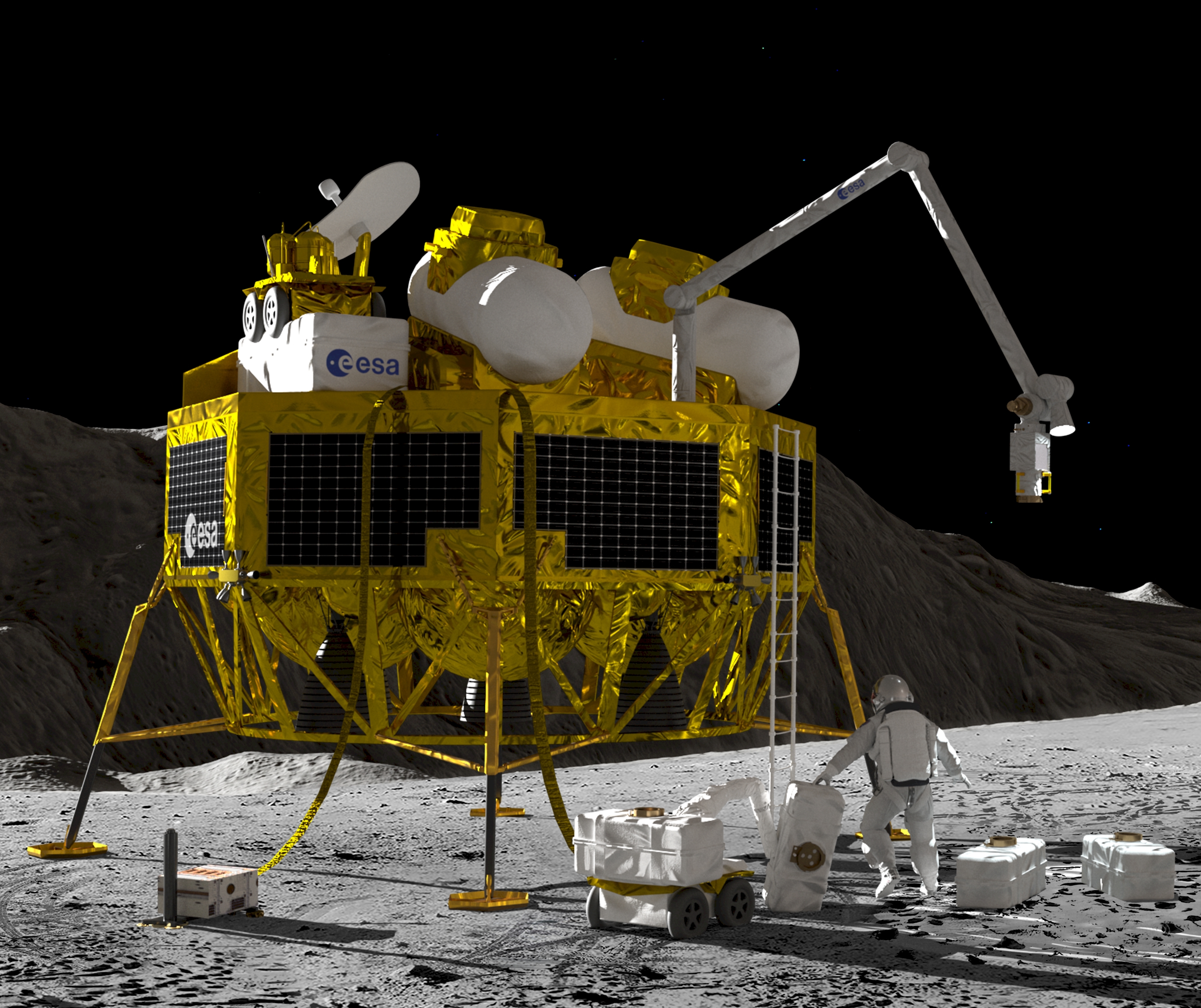}
    \caption{Artist's view of the Argonaut lander}
\end{wrapfigure}

Among the possible alternatives, Julia \cite{jl1} emerged as a particularly attractive candidate. Julia is a relatively recent programming language designed specifically for scientific computing, with the ambition of combining the ease of use of high-level interpreted languages with the performance traditionally associated with compiled languages such as C and C++. Its syntax is familiar to users of MATLAB and Python, making it easy for GNC engineers to adopt. Its performance claims \cite{jnasa} were the primary motivation for selecting Julia, as simulation speed is often a key driver in GNC development, directly impacting the feasibility of large-scale trade-off studies, robustness analyses, and verification campaigns.

This paper presents the authors' experience in developing ATLAS, a closed-loop Argonaut simulation framework written entirely in Julia. More specifically, it describes the transition from established MATLAB/Simulink-based development practices towards a software-centric Julia architecture, highlighting the challenges encountered and the lessons learned throughout the process. The objective is not only to assess the achievable simulation speed, but also to provide a practical evaluation of the language from the perspective of GNC engineers. The paper discusses the advantages offered by Julia, as well as its limitations, covering aspects such as software architecture, development workflow, debugging, performance optimization, and overall usability for closed-loop GNC simulations and analyses. 

\section{ATLAS architecture}

\subsection{GNC and simulation architecture}
%\begin{figure}
%    \centering
%    \includegraphics[width=0.4\linewidth]{images/arg.jpg}
%    \caption{Artist's view of the Argonaut lunar lander}
%    \label{fig:arg}
%\end{figure}

\begin{figure}
    \centering
    \includegraphics[width=0.9\textwidth]{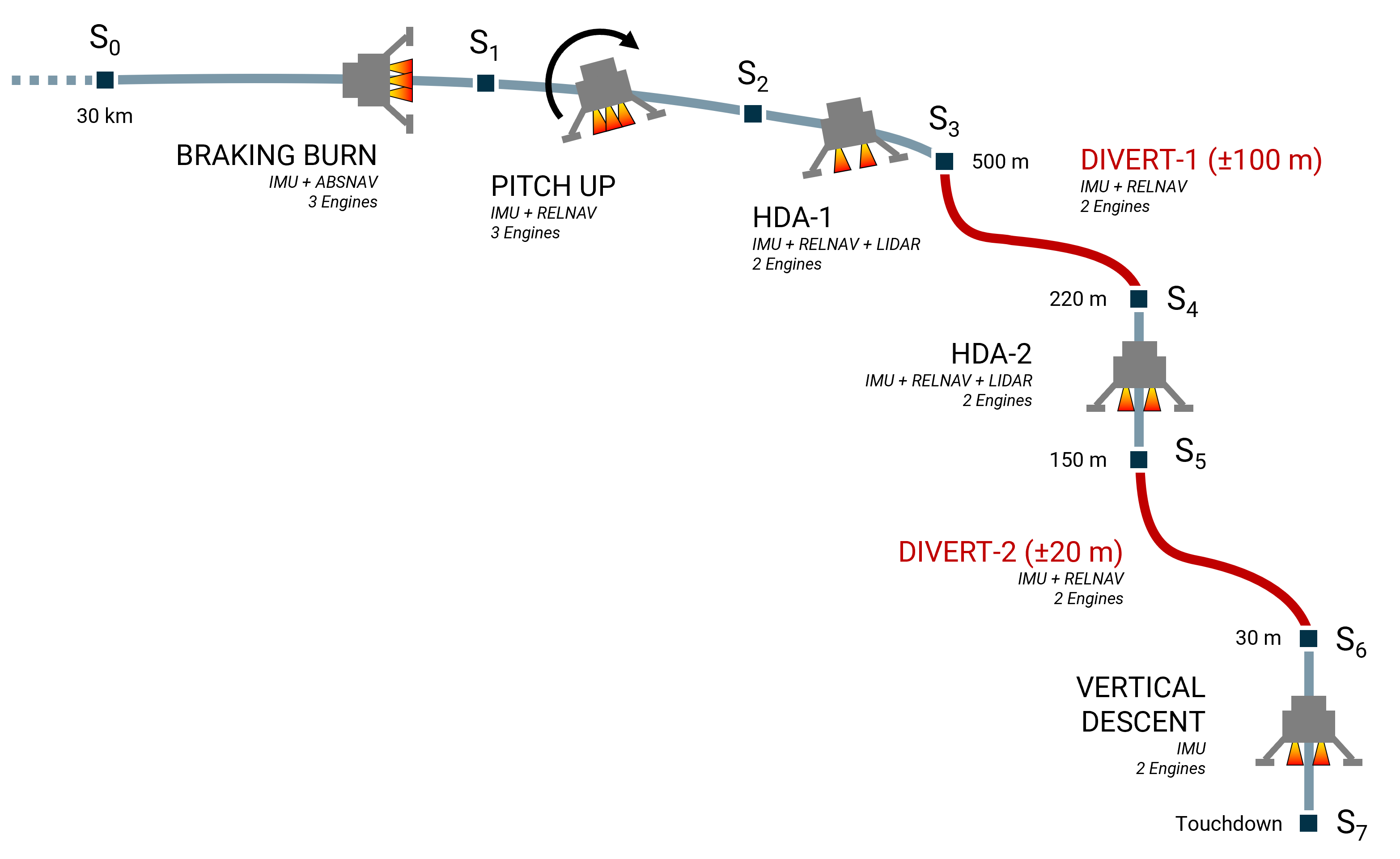}
    \caption{Argonaut descent and landing CONOPS}
    \label{traj}
\end{figure}

ATLAS focus is the Argonaut descent and landing phase, represented in Figure \ref{traj}, which starts 30 km above the lunar surface at the periselene of a 100$\times$30 km Elliptic Lunar Orbit (ELO). Argonaut trajectory reproduces the common scheme usually followed for lunar landings since the Apollo era \cite{b3}, comprising four main phases: a braking burn phase where the maximum thrust of main engines is used to significantly reduce the horizontal velocity of the vehicle; a pitch-up phase during which the thrust vector is reoriented to an almost vertical direction, allowing for the use of dedicated landing cameras and sensors; a powered descent for the last few hundreds of meters to reach a desired state above the landing site; and finally a vertical descent of few tens of meters at a given constant vertical speed to ensure a soft landing. The current engines' staging strategy foresees the use of three engines from waypoint $S_0$ to waypoint $S_2$ and two engines from $S_2$ to, while a Reaction Control System (RCS) comprising 24 thrusters provide pure torque and force control capability at all times. Argonaut Concept of Operations (CONOPS) can include up to two divert manoeuvres that can be autonomously triggered during the powered descent phase. The first one is located at an altitude of 500 m, and the second one is located above the designated landing site, at an altitude of 150 m. The combined divert manoeuvres can shift the horizontal position of the landing site to up to $\pm$ 120 m. 

\begin{figure}
    \centering
    \includegraphics[width=0.8\textwidth]{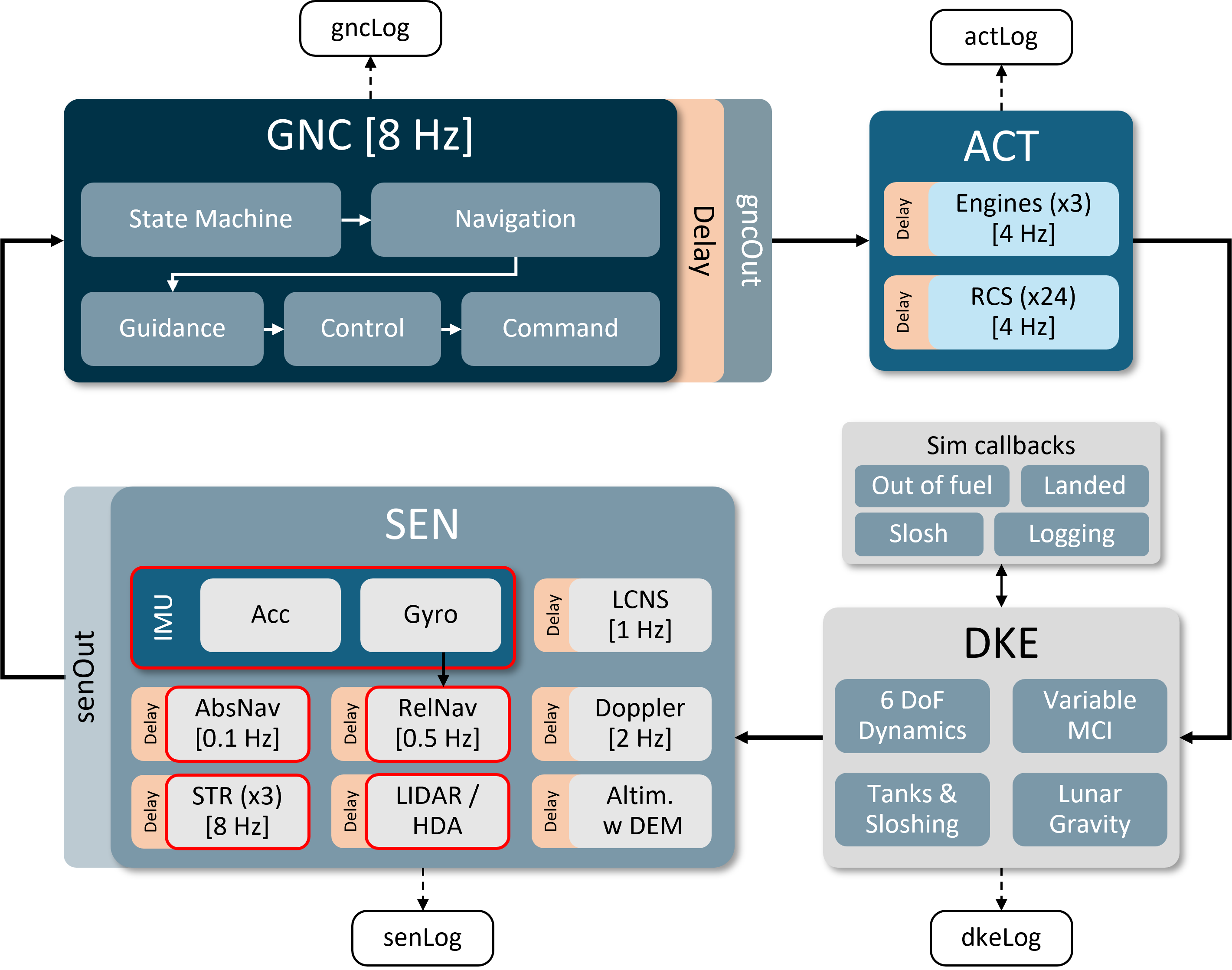}
    \caption{ATLAS simulation and software architecture. Baseline sensors for the descent and landing are highlighted in red.}
    \label{simArchitecture}
\end{figure}

Figure \ref{simArchitecture} illustrates the architecture of the ATLAS closed-loop simulation framework. The simulator follows a modular design, with a clear separation between the flight software (\texttt{GNC}), the sensor suite (\texttt{SEN}), the actuator chain (\texttt{ACT}), and the dynamics, kinematics, and environment models (\texttt{DKE}). These elements are interconnected through a closed feedback loop, including realistic timing, execution rates, and delays.

The \texttt{GNC} module is composed of five main functional blocks. A mission and vehicle manager implements the system state machine, governing mission phases, vehicle configurations, active units, and transitions between trajectory legs. The navigation function is centered on a six-degrees-of-freedom (6-DoF) error-state Schmidt--Kalman filter that fuses sensor measurements and IMU data to estimate the spacecraft inertial state. Additional navigation components provide onboard estimation of the vehicle mass properties and transform the estimated state into a landing-site-centered reference frame for use by the guidance and control functions. The guidance module generates the reference trajectory, including position, velocity, attitude, and angular-rate profiles, together with feed-forward main-engine thrust and RCS torque commands. Different guidance algorithms are employed for the various mission phases, following the design presented in \cite{b2}. The control architecture consists of four SISO control channels: one for vertical translation, one for roll, and two lateral channels controlling the coupled attitude-position horizontal motion. The lateral channels adopt a cascaded structure in which the commanded lateral forces generated by the outer position loops are converted into attitude commands, allowing the vehicle to redirect the main-engine thrust vector and achieve the desired translational motion. Finally, the command module translates the control demands into actuator-level commands, including main-engine thrust levels and RCS on-times, which are generated through a simplex-based allocation algorithm inspired by \cite{splx}.

The actuator module (\texttt{ACT}) contains high-fidelity models of the three main engines and the RCS. The main-engine model captures key implementation effects, including throttle and throttle-rate limitations, engine switch-on and switch-off delays, thrust realization errors, and detailed propellant consumption and gauging. Similarly, the RCS model incorporates realistic thruster dynamics and supports two modulation schemes. The first is Pulse Width Modulation (PWM), which closely simulates the physical operation of on-off thrusters. The second is Pulse Amplitude Modulation (PAM), in which the commanded firing time is converted into an equivalent average thrust level, providing a computationally efficient approximation of the PWM behavior. Both propulsion systems are commanded at a frequency of 4 Hz, consistent with the avionics and propulsion systems capabilities. However, the resulting thrust profiles are propagated continuously within the dynamics model, allowing the simulator to accurately capture the effects of actuator transients, delays, and modulation schemes on vehicle motion.

The dynamics, kinematics, and environment module (\texttt{DKE}) provides a high-fidelity 6-DoF representation of the vehicle. The model includes realistic propellant depletion, time-varying mass properties, and equivalent mechanical models of propellant sloshing for the four main-engine tanks, following the formulation presented in \cite{b1}. %Dedicated callback functions are employed to monitor mission-critical events and terminate the simulation when predefined conditions are met. These include nominal touchdown, propellant depletion, and the onset of sloshing instabilities, thereby preventing the simulation from propagating beyond physically meaningful operating conditions.

Finally, the sensor module (\texttt{SEN}) contains representative models of the Argonaut baseline navigation sensor suite, including Relative Vision-Based Navigation (RelNav), Absolute Vision-Based Navigation (AbsNav), inertial measurement units (IMUs), star trackers, and the LIDAR-based Hazard Detection and Avoidance (HDA) function. Additional sensors models are also included to support trade-off studies and future mission concepts, such as a Lunar Communication and Navigation Services (LCNS) receiver \cite{lcns}, a multi-beam Doppler radar, and a radar altimeter. The AbsNav function is implemented as a performance model representative of crater-based absolute navigation algorithms \cite{crat}\cite{dlr}. Synthetic crater catalogs are generated and projected onto the camera image plane according to the spacecraft attitude and position, from which crater-bearing measurements are derived and provided to the navigation filter. Similarly, the RelNav function includes the generation and tracking of synthetic opportunistic surface features, together with a complete two-point visual odometry pipeline inspired by \cite{vo}. This module provides relative motion measurements between selected key-frames, enabling robust estimation of the vehicle direction of motion during the terminal descent phase. To complement the navigation suite, the HDA model emulates the landing site selection process by randomly selecting a safe landing location within the accessible LIDAR field of view. The resulting (random) target location is then provided to the guidance system, allowing the simulation of divert maneuvers and the associated navigation, guidance, and control interactions in a closed-loop environment.

Overall, the proposed architecture simulates a realistic closed-loop implementation: GNC computes commands based on delayed, noisy sensor inputs; actuators apply those commands with their own dynamics and delays; the DKE propagates the true physical state; and the sensors observe that state imperfectly and feed it back. The inclusion of asynchronous and multi-rates modules, delays, and detailed units modelling makes the setup suitable for high-fidelity performance assessment of guidance, navigation, and control architectures under realistic operational conditions.

\begin{figure}
\centering
\begin{minipage}{0.8\textwidth}
%\begin{tcolorbox}[mycode]
%\begin{minted}[
%    bgcolor=codebg,
%    fontsize=\footnotesize,
%    breaklines,
%    tabsize=4,
%]{julia}
\begin{tcblisting}{mycode}
# Dynamics,kinematics, and environment struct
struct DKE
    # Model parameters
    mu::Float64                     # [m3/s2] Moon gravitational parameter
    R::Float64                      # [m] Moon radius
    <...>

    # Sub-models structs
    mci::MCI                        # Mass-centering-inertia of the vehicle
    moonGravity::GravityField       # Moon gravity field
    <...>

    # Allocation/post-processing variables
    torque_B::Vector{Float64}       # [Nm] Torque applied on the vehicle
    force_B::Vector{Float64}        # [N] Force applied on the vehicle
    <...>
end

# Log variables for post-processing
@inline function logDke(t, x, dke::DKE)
    return (
        torque_B=copy(dke.torque_B),
        force_B=copy(dke.force_B),
        <...>
    )
end
\end{tcblisting}
%\end{minted}
%\end{tcolorbox}
\end{minipage}
\caption{Implementation of DKE structure.}
\label{dkeStruct}
\end{figure}

\begin{figure}
\centering
\begin{minipage}{0.8\textwidth}
%\begin{tcolorbox}[mycode]
%\begin{minted}[
%    bgcolor=codebg,
%    fontsize=\footnotesize,
%    breaklines,
%    tabsize=4,
%]{julia}
\begin{tcblisting}{mycode}
struct SIM_PARAMS
    dke::DKE
    act::ACT
    sen::SEN
    gnc::GNC
end

# This function handles the call to sensors, GNC, and any discrete-time part 
# of actuators, at time t[k]. Delays between the different modules are handled 
# by the modules themselves.
function simStep!(integrator)
    t, x, p = integrator.t, integrator.u, integrator.p  # t[k], x[k], p[k]
    outDke!(p.dke, t, x)
    senOut = executeSen!(t, x, p.dke, p.sen, p.gnc)
    gncOut = executeGnc!(t, p.gnc, senOut)
    executeAct!(t, p.act, gncOut)
end

@inline simLog(x, t, integrator) = (
    gncLog=logGnc(integrator.p.gnc),
    dkeLog=logDke(t, x, integrator.p.dke),
    senLog=logSen(integrator.p.sen),
    actLog=logAct(t, integrator.p.act),
)
\end{tcblisting}
%\end{minted}
%\end{tcolorbox}
\end{minipage}
\caption{Implementation of simulation parameters, simulation step function, and logging.}
\label{simStep}
\end{figure}

% (for example,sbehavior(ts) ()modelingbehavior DKE()mentionsuggestsermay sensorsormodeling
\subsection{Software architecture}
ATLAS has been designed from the ground up with modularity as a fundamental architectural principle, and with \texttt{DifferentialEquations.jl} \cite{rack} as dynamics propagation backbone. The objective is to decompose the simulation into self-contained components that can be developed, tested, and understood independently, while still forming a coherent hierarchical system when assembled together. This approach promotes maintainability, reusability, and extensibility, which are essential characteristics for a simulation framework intended to evolve alongside the development of a complex spacecraft project. The implementation relies heavily on Julia's composite types (\texttt{struct}), which provide a natural mechanism to represent engineering systems composed of multiple interacting subsystems. Through the recursive composition of structures within structures, complex models can be represented hierarchically, mirroring the physical decomposition of the vehicle, its onboard functions, and units.

Each of the four top-level modules shown in Figure \ref{simArchitecture}, namely \texttt{GNC}, \texttt{ACT}, \texttt{DKE}, and \texttt{SEN}, follows a common software architecture based on custom Julia \texttt{struct}s. Figure \ref{dkeStruct} illustrates this design using the DKE module as an example. Each structure groups together the model parameters, the relevant sub-model structures, and a set of cache variables used during simulation execution. These caches serve two main purposes. First, they store intermediate quantities that are repeatedly accessed throughout the simulation but do not constitute model states or outputs. Second, they provide preallocated memory buffers that significantly reduce heap allocations during runtime, which is a key aspect of achieving high simulation performance in Julia. In addition, every module implements a dedicated logging interface responsible for collecting the variables required for post-processing and analysis without polluting the core computational logic.

The execution of the closed-loop simulation is orchestrated through periodic calls of \texttt{simStep!}, whose code is reported in Figure \ref{simStep}. This function is implemented as a \texttt{PeriodicCallback} and acts as the discrete-time logical element of the simulator. Its primary input is the global \texttt{SIM\_PARAMS} structure, which contains references to all major simulation modules and provides a unified interface to the complete simulator state. At each callback invocation, the outputs of the various subsystems are generated sequentially, ensuring a deterministic execution order throughout the closed-loop chain. The callback period corresponds to a fundamental simulation timestep chosen as the greatest common divisor of all module execution rates and timing delays. ATLAS design also enables the coexistence of components operating at different frequencies while maintaining a single synchronized simulation framework. To support this functionality, each module incorporates an internal triggering and delay management mechanism. Trigger counters determine when a given function shall be executed, while dedicated First-In-First-Out (FIFO) buffers model processing delays and data latencies. The output generated by a module can therefore be stored and retrieved at a later time, allowing realistic representation of sensor processing times, command latencies, and flight-software scheduling effects. Figure \ref{gnccall} illustrates how this mechanism is implemented in the \texttt{GNC} module.

In addition to \texttt{simStep!}, the simulator relies on several scenario-dependent \texttt{ContinuousCallback}s to monitor the evolution of the mission and detect critical events. These callbacks automatically terminate the numerical integration when the vehicle reaches touchdown, exhausts its available propellant, or violates predefined sloshing stability limits. This event-driven architecture avoids propagating the simulation beyond physically meaningful operating conditions while improving computational efficiency. Finally, a dedicated \texttt{SavingCallback} is responsible for collecting the variables exposed by the individual module logging interfaces and storing them for subsequent post-processing. This separation between simulation execution and data collection further contributes to the modularity of the framework and allows logging needs to evolve independently of the underlying simulation models.

\begin{figure}
\centering
\begin{minipage}{0.53\textwidth}
%\begin{tcolorbox}[mycode]
%\begin{minted}[
%    bgcolor=codebg,
%    fontsize=\footnotesize,
%    breaklines,
%    tabsize=4,
%]{julia}
\begin{tcblisting}{mycode}
function executeGnc!(t, gnc::GNC, senOut)
    gncOut = GNC_OUT_EMPTY

    if triggerNow(gnc.del, t)
        gncMode = executeModeManager!(gnc, senOut)
        executeNavigation!(gnc, senOut)
        executeGuidance!(gncMode, gnc)
        executeControl!(gncMode, gnc)
        gncOut = executeCommand!(gncMode, gnc)
    end

    return getDelayedOutput!(gnc.del, t, gncOut)
end
\end{tcblisting}
%\end{minted}
%\end{tcolorbox}
\end{minipage}
\caption{GNC execution function.}
\label{gnccall}
\end{figure}

\section{Julia advantages and limitations}
This section summarizes the main lessons learned from the development of ATLAS, viewed from the perspective of GNC engineers with extensive experience in MATLAB/Simulink-based simulation and algorithm development. Rather than providing a comprehensive review of the Julia ecosystem, the objective is to discuss the practical advantages and limitations encountered while designing, implementing, and using a high-fidelity closed-loop simulator for a real aerospace application. 

It should be noted that the ATLAS development effort started several years ago. Since then, Julia and its ecosystem have continued to evolve rapidly, with new packages, capabilities, and performance improvements being introduced on a regular basis. Consequently, some of the observations presented in this section may not fully reflect the latest developments available at the time of publication. Nevertheless, they remain representative of the experience gained throughout the development of ATLAS and of the challenges encountered when transitioning from established MATLAB/Simulink workflows to a Julia-based simulation framework. 

Finally, this discussion is intentionally restricted to license-free, open-source packages. Commercial products and Julia-based industrial platforms, such as Dyad, are therefore outside the scope of this work. The aim is to assess the capabilities and limitations of a purely open-source Julia ecosystem and to evaluate its suitability as an alternative to traditional MATLAB/Simulink-based approaches for GNC simulation and analysis.

\subsection{Ecosystem maturity, licensing, portability}
Matlab has long established itself as the de facto industrial standard for GNC software prototyping and development, largely due to the consistency, reliability, and depth of its ecosystem. The toolchains are tightly integrated, extensively tested, and supported by a professional development infrastructure that ensures long-term stability. This robustness is reinforced by a rapid release cycle and strong backward compatibility practices, which allow large engineering organizations to build and maintain complex simulation frameworks over many years with relatively low risk.
However, these advantages come with structural limitations. Licensing costs are significant, and in large organizations access is often managed through floating license pools. This introduces friction at the operational level: engineers may not always have immediate access to the tools they need, especially during peak usage periods. As a result, even well-designed simulation frameworks can become less accessible in practice, simply due to licensing constraints rather than technical limitations. 

Julia, by contrast, offers a fundamentally different paradigm. It is open source, free of licensing constraints, and designed from the ground up for high-performance numerical computing. Its syntax is expressive and close to mathematical notation, making it appealing for modelling and algorithm development. While the ecosystem has grown rapidly, it still exhibits a certain heterogeneity in terms of maintenance, code and documentation quality, and long-term support. Many packages originate from academic environments or small developer teams, which can result in uneven levels of testing rigour and sustainability. Consequently, some libraries may become inactive, posing potential risks for long-term projects. This stands in contrast to Matlab’s centrally managed and professionally maintained toolbox ecosystem. Nevertheless, Julia includes several outstanding packages that represent the state of the art in numerical computing. A prominent example — heavily used within ATLAS — is the \texttt{DifferentialEquations.jl} suite \cite{rack}, which is widely recognized for its advanced numerical integration capabilities, often delivering both superior performance and flexibility compared to alternative tools. Comparable high-quality solutions are also available in areas such as optimization, linear algebra, and plotting \cite{makie}.

The key shift introduced by the use of an open-source tool as Julia for GNC design is not only technical, but organizational. A system engineer can clone a repository, install Julia, and immediately run a high-fidelity simulation without needing to request licenses, configure complex environments, or depend on centralized IT infrastructure. This democratizes access to advanced simulation tools and enables a more distributed and flexible workflow. At the project level, this has important implications. Tools become inherently more portable and reproducible, facilitating collaboration across teams. It becomes easier to share models, validate results independently, and integrate simulation into broader development pipelines such as continuous integration or digital-twins modelling. In other words, the simulation ceases to be a specialized resource tied to a specific tooling environment and instead becomes a readily available asset that can be used, adapted, and extended to a wider range of stakeholders.

\subsection{Language semantics}
%Starting from zero
%Multiple dispatch
%Passed by references%dynamic typing (that feels like static typing) that promotes type stability

A source of difficulty when working with Julia lies in its language semantics, in particular the handling of data through references rather than copies. In Julia, most objects — especially arrays and other mutable structures — are passed by reference by default. This means that functions typically operate on the original data unless explicit steps are taken to create a copy. While this behaviour is consistent with many high-performance programming languages and avoids unnecessary memory allocations, it introduces a class of errors that can be subtle and difficult to diagnose.
In contrast, MATLAB adopts a copy-on-write strategy, whereby variables behave as if they were passed by value. From the user’s perspective, this means that modifying a variable inside a function does not affect the original data unless explicitly returned and reassigned. This greatly reduces the likelihood of unintended side effects and makes the behaviour of code more predictable, particularly for users who are not explicitly reasoning about memory and data ownership.
The difference becomes particularly relevant in closed-loop simulations, where the system state, control inputs, and intermediate variables are repeatedly updated within tightly coupled loops. In Julia, inadvertent aliasing—where multiple variables reference the same underlying data—can lead to silent corruption of the simulation state. Such errors often do not produce immediate failures, but instead manifest as incorrect simulation results, making them significantly harder to trace and debug.
As a consequence, Julia places a greater burden on the developer to be aware of when data is shared and when it is explicitly duplicated. This requires a more disciplined programming style and a deeper understanding of the language’s semantics. In practice, this can make Julia less forgiving than MATLAB, where the default behaviour prioritizes safety and ease of use over strict performance considerations. While Julia’s approach is ultimately more efficient and flexible, it also makes it much easier to introduce subtle bugs that are considerably more difficult to identify, particularly in large and complex simulation environments.

\subsection{Prototyping speed: coding approach}
Julia is often presented as a solution to the so‑called \emph{two-language problem}, namely the need to combine a high-level, user-friendly language for rapid development with a low-level, high-performance language for computational efficiency. In principle, Julia addresses this issue by allowing users to write expressive, MATLAB-like code while relying on just-in-time compilation and type specialization to achieve execution speeds comparable to C or C++.
This dual capability is fundamentally valid, but its practical realization requires careful development. Julia’s syntax and overall feel make it particularly well-suited for rapid prototyping: array-based operations and concise function definitions closely resemble the MATLAB programming model. This allows users to quickly implement simulation functions and models in a clear and compact manner. However, code written in a purely MATLAB-like style often relies on dynamically typed variables, implicit memory allocation, and generic container types. Such patterns inhibit the compiler’s ability to infer concrete types and optimize execution. As a result, performance may degrade significantly, in some cases even exceeding the execution speed of interpreted environments. 
Achieving performance comparable to C++ typically requires adopting a more disciplined programming style. In Julia, this involves ensuring type stability, using concrete data structures, preallocating memory, and favouring in-place operations. While these practices do not introduce the same level of verbosity or manual memory management found in C++, they nonetheless impose a structure that departs from the most straightforward MATLAB-like coding approach. To better illustrate this with a concrete example, Figures \ref{kalmannaive} and \ref{kalmanopt} show an example of naive and optimized implementation of a Kalman filter update function.

\begin{figure}[h!]
\centering
\begin{minipage}{0.56\textwidth}
%\begin{tcolorbox}[mycode]
%\begin{minted}[
%    bgcolor=codebg,
%    fontsize=\footnotesize,
%   breaklines,
%    tabsize=4,
%]{julia}
\begin{tcblisting}{mycode}
function kalmanNaive(x, P, y, yEst, R, H)
    K = (P * H') / (R + H * P * H')
    x = x + K * (y - yEst)
    P = (I - K * H) * P
    return x, P
end
\end{tcblisting}
%\end{minted}
%\end{tcolorbox}
\end{minipage}
\caption{Naive implementation of Kalman update}
\label{kalmannaive}
\end{figure}
\begin{figure}[h!]
\centering
\begin{minipage}{0.56\textwidth}
%\begin{tcolorbox}[mycode]
%\begin{minted}[
%    bgcolor=codebg,
%    fontsize=\footnotesize,
%    breaklines,
%    tabsize=4,
%]{julia}
\begin{tcblisting}{mycode}
function kalman!(nav::NavState, y, meas::NavMeas)

    # Extract pre-allocated data from 'nav' and 'meas'
    x = nav.x; P = nav.P; nx = nav.nx
    yEst = meas.yEst; PHt = meas.PHt; 
    S = meas.S; R = meas.R; K = meas.K; 
    H = meas.H; ny = meas.ny; dy = meas.dy

    # Measurement innovation
    mul!(PHt, P, H')
    mul!(S, H, PHt)        
    S .+= R
    dy .= y .- yEst

    # Compute Kalman Gain
    K .= PHt
    rdiv!(K, cholesky!(Hermitian(S)))

    # Update state and covariance matrix
    @inbounds for r in 1:nx, j in 1:ny
        k = K[r, j]
        x[r] += k * dy[j]
        for c in r:nx
            P[r, c] -= k * PHt[c, j]
            P[c, r] = P[r, c]
        end
    end

    return x, P
end
\end{tcblisting}
%\end{minted}
%\end{tcolorbox}
\end{minipage}
\caption{Optimized implementation of Kalman update}
\label{kalmanopt}
\end{figure}

\begin{table}[htbp]
  \centering
  \begin{tabular}{@{}lcccc@{}}
    %\toprule
    \textbf{Metric} & & \texttt{kalmanNaive} (Julia) & \texttt{kalmanNaive} (Matlab) & \texttt{kalman!} (Julia)\\
    \hline
    Time range [min, max] & $\mu$s  & [74.9, 2154.2]    & [9.077, \textbf{65.407}]  & [\textbf{3.388}, 193.387]\\
    Median time & $\mu$s            & 72.600            & 11.409                    & \textbf{3.600}\\
    Mean time & $\mu$s              & 100.335           & 12.216                    & \textbf{3.631}\\
    Std. deviation & $\mu$s         & 110.577           & 3.529                     & \textbf{2.111}\\
    Memory usage & KiB              & 65.52             & --                        & \textbf{0}\\
    Number of allocations & -       & 31                & --                        & \textbf{0}\\
    \hline\\
  \end{tabular}
  \caption{Comparison of naive (Julia and Matlab) and optimized (Julia only) Kalman update implementations (50 states, 3 scalar measurements) over 10,000 function calls.}
  \label{tab:kalman_impl}
\end{table}

Table \ref{tab:kalman_impl} indicates the execution time difference that can be seen between the naive implementation of the Kalman update and a Julia-optimized version. Beyond simply running (consistently) faster, an optimized approach can also help with reducing the code's memory footprint, by minimizing memory allocations and making use of existing variables wherever possible. We note that the optimized function \texttt{kalman!} performs significantly better than a comparable Matlab implementation, while Matlab beats the naive Julia implementation. %A faithful complement to the optimized code version is not possible in Matlab as it lacks distinct features that prevent an accurate implementation. Namely these include: In-place allocation of a matrix multiplication, in-place allocation of a Cholesky decomposition, and the lack of a comparable \texttt{@inbounds} macro. Interpreted scalar loops are possible but quite slow and thus unsuitable for performance-oriented code. 

Julia’s key advantage lies in the fact that expressiveness and speed can be addressed within a single language framework. Unlike traditional workflows, where performance-critical components are (re-)written in C or C++ after prototyping, Julia allows developers to incrementally refine the same codebase. A simulation can initially be written in a high-level, MATLAB-like style to validate algorithms and system behaviour. Subsequently, profiling tools can be used to identify bottlenecks, and only the relevant sections need to be refactored to adopt performance-oriented patterns. This process preserves code continuity while progressively improving execution efficiency.
In this sense, Julia does not eliminate the tension between ease of use and performance, but rather relocates it within the language itself. The user must still navigate a spectrum between readability and efficiency; however, this trade-off is managed through coding style rather than through a transition between different programming languages. It is the opinion of the Authors that a more precise characterization, therefore, is that Julia provides the capability to unify high-level expressiveness and low-level performance, but realizing both simultaneously is generally not easy or even possible.

\subsection{Prototyping speed: modelling IDE}
Another important aspect when adopting Julia for closed-loop simulations is the absence of a widely used, free, fully integrated graphical modelling environment comparable to Simulink. Consequently, simulations in Julia are typically developed as conventional software, written directly in code rather than assembled through graphical blocks.

This represents a significantly higher barrier to entry, particularly for users accustomed to model-based design tools. In a Simulink-like environment, system dynamics, control laws, and signal flows can be constructed visually, allowing for rapid iteration with minimal concern for implementation details such as data structures, memory management, or execution order. By contrast, Julia requires the user to explicitly structure the simulation, define interfaces between components, and manage the flow of data through the system. This makes the initial development process slower and demands stronger software engineering skills, especially when building complex closed-loop architectures involving multi-rate and delayed functional elements.
However, this apparent disadvantage is accompanied by a number of important benefits, particularly in the context of guidance, navigation, and control applications. Writing simulations as pure software naturally enforces a more disciplined and explicit representation of system behaviour. Interfaces between components must be clearly defined, data types must be consistent, and execution logic must be unambiguous. This tends to produce implementations that are closer in structure to actual flight software, where determinism, traceability, and maintainability are critical.

In contrast, graphical modelling environments such as Simulink can, over time, lead to increasingly complex and difficult-to-maintain models. Large block diagrams may evolve into so-called “spaghetti models”, where signal routing, hidden dependencies, and implicit execution order obscure the underlying logic. While tools like Simulink excel at rapid prototyping and visualization, scaling them to large, high-fidelity closed-loop simulations can introduce challenges in terms of readability, task execution, functional logic, and data transfer verification.

By forcing the user to operate within a code-based paradigm from the outset, Julia encourages the development of modular, testable, and maintainable simulation components. This reduces the conceptual and structural gap between simulation and implementation, potentially lowering the effort required for model-to-flight-code transition.
In this sense, while the lack of a Simulink-like environment makes Julia a more demanding starting point, it promotes better GNC software architecture and design. The resulting simulation frameworks may require more upfront effort to develop, but they tend to be more robust, more transparent, and more closely aligned with operational GNC software design.

\subsection{Misses and gaps}
A major practical limitation of Julia, particularly evident in interactive development workflows, is the so‑called \emph{Time To First X} (TTFX). This refers to the latency introduced by just-in-time compilation when code is executed for the first time. Even for relatively simple simulations, the initial run can incur noticeable delays as functions, dependencies, and type specializations are compiled. This effect is not limited to the very first execution of a script, but reappears whenever significant code modifications are introduced, invalidating previously compiled methods and triggering recompilation.

In the context of GNC design and simulations, this behaviour can become a substantial impediment to productivity. Development in this domain is inherently iterative, involving frequent adjustments to models, parameters, and algorithms. Each such change may lead to renewed compilation overhead, interrupting the natural edit–run–analyse cycle. This introduces friction that is both perceptible and cumulative over time.
While various strategies exist to mitigate TTFX — such as function precompilation or the use of system images — they require additional effort and expertise, and often do not entirely eliminate the issue. As a result, Julia can feel significantly less responsive during early-stage development and experimentation. This stands in contrast to its excellent runtime performance once execution is underway, highlighting a trade-off between compilation latency and execution speed that remains a notable gap in practice.

A further limitation lies in the debugging experience, which remains less intuitive compared to more interactive environments such as MATLAB and Simulink. In Simulink, debugging is tightly integrated into the graphical modelling paradigm, allowing users to inspect signals with scopes, trace data flows, and step through execution in a highly intuitive manner. Breakpoints can be placed directly within block diagrams, and the evolution of system variables can be monitored visually in real time. This makes it particularly effective for diagnosing issues in closed-loop systems, where understanding the interaction between components is essential.
In contrast, debugging in Julia is more traditional and code-centric, and although functional, it lacks the same level of integration and immediacy. Interactive debugging can feel less fluid, especially when dealing with complex call stacks or type-related errors. Moreover, error messages—while informative—can become difficult to interpret in situations involving multiple dispatch or type inference failures, which are common in performance-oriented Julia code. These factors can slow down the identification and resolution of issues, particularly in large simulation environments where faults may propagate across several layers of abstraction.

\subsection{Simulation speed}
Simulation speed is the primary motivation for adopting Julia in the first place, particularly for GNC applications where large Monte Carlo campaigns with high-fidelity models can quickly become computationally prohibitive. 

\begin{table}[!h]
    \centering
    \begin{tabular}{ccc}
    & \textbf{Processor} & \textbf{RAM}\\ \hline
    \textit{Laptop} & Intel Core i7-12800H (2.40 GHz) & 32 GB \\
    \textit{Workstation} & Intel Core i9-13900K (3.00 GHz) &  128 GB\\ \hline
    & 
    \end{tabular}
    \caption{Computing stations used for the assessment of the simulation speed.}
    \label{cmpt}
\end{table}

\begin{figure}
    \centering
    \includegraphics[scale=0.6]{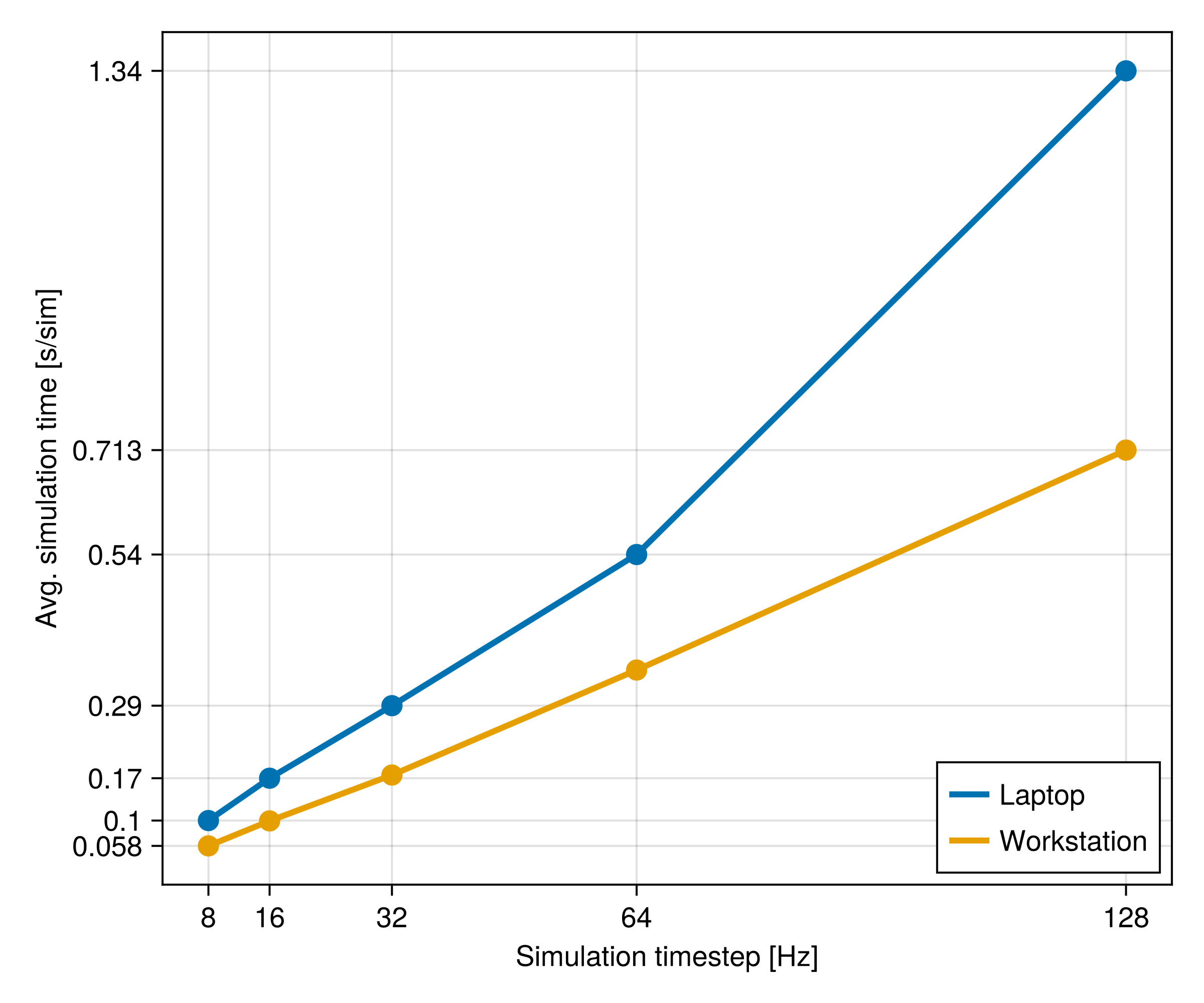}
    \caption{Average time required to simulate one trajectory in a Monte-Carlo campaign of 100 simulations.}
    \label{simtime}
\end{figure}

%The simulated Argonaut scenario focuses on the final 900 seconds of the descent and landing trajectory prior to touchdown. The nominal GNC baseline was modelled throughout this phase, including Absolute \& Relative Vision-Based Navigation and IMU as sensing chains, with detailed performance models for the image-processing functions, and the complete GNC algorithms running in closed-loop. 
To assess simulation speed, a Monte Carlo campaign of 100 descent and landing trajectories was conducted with ATLAS, using the two different computing stations described in Table \ref{cmpt}. For each platform, the campaign was repeated across a range of fixed simulation timesteps (i.e., fixed ODE integration steps), using the default \texttt{Tsit5()} solver \cite{tsit} from \texttt{DifferentialEquations.jl} configured in fixed-step mode. Each Monte Carlo simulation was executed in parallel using multithreading via \texttt{EnsembleThreads}, with Julia launched using the \texttt{--threads=auto -O3} configuration. For every setup, the average wall-clock time required to complete a single trajectory (with a simulated time of flight of approximately 900 s) was measured. The results are summarized in Figure \ref{simtime}. 

At a simulation rate of 8 Hz (i.e., one integration step per GNC cycle), a single trajectory can be simulated in approximately 100 ms on the laptop and less than 60 ms on the workstation. At these speeds, a typical Monte Carlo campaign of 1000 simulations can be executed in under 2 minutes. Even when increasing the simulation rate to a more representative value of 16 Hz, execution times remain below 0.2 seconds per trajectory, corresponding to less than 3.5 minutes for a 1000-run campaign. Unfortunately, no direct comparison with an equivalent MATLAB/Simulink-based simulator was performed, as this would have required the development of a fully independent reference implementation. Nevertheless, based on the Authors’ experience with comparable Phase B simulators, it is reasonable to expect that the execution times reported here are lower than those typically achieved by expert MATLAB developers within a plain MATLAB/Simulink environment (i.e., not exploiting MATLAB's Parallel Computing Toolbox and Coder -- both needing additional specific licenses). 

We clearly see that Julia is capable of delivering performance comparable to low-level languages, but achieving this in practice requires a non-negligible amount of \emph{software engineering} effort. Two aspects are especially critical when trying to maximize simulation speed. First, it is essential to leverage state-of-the-art libraries that are specifically designed for performance. In the context of dynamical system simulation, packages such as the aforementioned \texttt{DifferentialEquations.jl} provide highly optimized and extensively tested solvers that, when used correctly, can significantly outperform hand-written integration loops. %These libraries encapsulate advanced numerical methods, adaptive step-sizing strategies, and efficient internal representations, allowing users to benefit from sophisticated algorithms without reimplementing them from scratch.
Second, and arguably more important, is the need to minimize heap allocations. Memory allocation is a major source of performance degradation in Julia, particularly in high-frequency simulation loops where small inefficiencies are amplified over many iterations. Unintended allocations can arise from seemingly innocuous coding patterns that are common in other programming languages, such as creating temporary arrays or using non type-stable operations. 

ATLAS results were not obtained from the outset, but rather through a progressive and effort-intensive optimization process. During the early stages of the development, a more naive coding approach was adopted, closely resembling MATLAB-style practices and not fully leveraging Julia’s performance. In this initial phase, execution times exceeded 5 seconds per simulation, highlighting that simply using Julia does not automatically yield high performance.
As the implementation matured, incremental improvements -- most notably the introduction of multithreading -- enabled a reduction in execution time to approximately one simulation per second. While this represented a significant improvement, it remained far from the performance ultimately achieved and expected. A major turning point was a comprehensive refactoring phase aimed specifically at reducing and, where possible, eliminating heap allocations within the simulation loop. This required substantial restructuring of the codebase, including the adoption of in-place operations, improved type stability, and more efficient data handling patterns. This optimization effort proved decisive in unlocking Julia’s full performance potential, ultimately reducing execution time to approximately 0.1 seconds per simulation. 
Overall, in the ATLAS development process, the design and prototyping of the GNC functions and of simulation models took roughly 20\% of the effort; developing the simulation framework that would mimic Simulink capabilities took approximately another 30\% of development effort. The remaining 50\% was put in reducing allocations during simulations, which required explicit control over data structures, preallocation of buffers, and the ubiquitous use of in-place operations. 

\subsection{New ways of working}
Beyond the technical merits of the language itself, the most significant benefit observed during the development of ATLAS was the emergence of new and more efficient GNC engineering workflows. Historically, simulation execution time has often been a limiting factor in design and verification activities. When a single Monte Carlo campaign requires hours or days to complete, engineers are naturally forced to be selective in the analyses they perform. In contrast, the execution speeds achieved with ATLAS have made large-scale simulations a routinely accessible tool, allowing simulation results to influence design decisions much earlier and more frequently throughout the development process.

A representative example is the tuning of Argonaut control system. Traditional workflows typically rely on linear analysis techniques, such as frequency-domain assessments and linearized models, to design and tune controllers. While these methods remain essential, they must be complemented by nonlinear closed-loop simulations to validate the resulting design. With ATLAS, the computational cost of these simulations became sufficiently small to allow rapid iterations between controller design and Monte Carlo validation. Candidate tuning parameters could be evaluated through hundreds of nonlinear simulations, the results analyzed, and the controller retuned almost interactively. This enabled a much tighter integration between classical control design techniques and nonlinear performance validation than is typically feasible with slower simulation environments.

The same applies to GNC verification activities. Demonstrating compliance with performance requirements at the typical probability and confidence level required for class-$\alpha$ space missions generally requires the execution of multiple Monte Carlo campaigns with thousands of simulations each. Such campaigns are often among the most time-consuming activities in the verification and validation process. The simulation time achieved with ATLAS suggests a future in which the computational burden of these analyses is no longer a primary constraint. Instead, engineers can focus on defining representative uncertainty models and meaningful performance metrics, rather than limiting the scope of the analysis due to available computing or time resources. This has the potential to substantially reduce the effort associated with verification campaigns and, ultimately, shorten GNC development schedules by weeks or even months.

Furthermore, the results presented in this paper were obtained using conventional multicore CPUs only. Additional performance gains can reasonably be expected through the adoption of emerging GPU-based frameworks such as \texttt{DiffEqGPU.jl} and \texttt{Reactant.jl}, which aim to accelerate large ensembles of simulations and computationally intensive numerical workloads. The Julia ecosystem is currently experiencing rapid growth in this area, with GPU support becoming increasingly mature and accessible. As these technologies continue to evolve, the gap between simulation execution time and engineering decision-making is expected to narrow even further, opening the door to design and verification methodologies that would previously have been considered computationally impossible.

\subsection{An industrial perspective}
While the results presented in this paper are encouraging from a simulation and analysis perspective, the ultimate objective of GNC development is not to produce a simulator but rather a verified flight software application compliant with ECSS software engineering standards and, where applicable, safety requirements. Consequently, the value of any development framework must ultimately be assessed not only in terms of simulation performance, but also in terms of its compatibility with industrial flight-software development processes. Over the last decade, the European space industry has largely converged towards model-based development practices centred on MATLAB/Simulink and automatic code generation \cite{savoir1}\cite{savoir2}. These workflows have matured considerably and are now deeply embedded in industrial processes, toolchains, and qualification activities. As a result, a complete ecosystem exists today to support the transition from GNC algorithms implemented in Simulink to flight software.% while maintaining traceability, verification evidence, and compliance with applicable standards. 

In this context, a key open question concerns the role that Julia could play beyond simulation and GNC prototyping. Although some solutions start to emerge within the open-source ecosystem to generate deployable code from Julia applications, their maturity for spacecraft applications remains, to the best knowledge of the Authors, largely unproven. More importantly, it is unclear how such approaches could be integrated into industrial development and qualification processes while satisfying the stringent requirements imposed by ECSS standards and, for safety-critical applications, by software assurance activities. The technical feasibility of generating onboard code may therefore be only one part of the challenge, as the significant investments already made by industry in Simulink-based autocoding workflows make a rapid shift towards a completely different paradigm unlikely. 

Consequently, the most realistic near-term industrial application of Julia may be in the pre-development phases of a programme, including feasibility studies, Phase A/B activities, trade-off analyses, architecture assessments, and early GNC prototyping. In such a scenario, Julia could provide substantial benefits in terms of simulation performance and development flexibility, while the flight-software development would ultimately transition to a more traditional, qualification-ready toolchain. Whether Julia can evolve beyond this supporting role and become part of the operational flight-software development process remains an open question that will require both technological and industrial maturation.

\section{Conclusions}
This paper presented ATLAS, a high-fidelity closed-loop simulation framework for the ESA Argonaut lander developed entirely in Julia. The framework was created to support independent GNC assessment activities within ESA while simultaneously exploring an alternative to the traditional MATLAB/Simulink-based simulation paradigm. The results demonstrate that Julia can deliver excellent runtime performance for large-scale GNC analyses. Through the use of optimized numerical libraries, multithreading, and extensive reduction of heap allocations, ATLAS achieved execution times of approximately 0.1 seconds for a complete 900-second landing simulation. At the same time, the development experience highlighted several challenges associated with the language. Achieving high performance required significant software engineering effort, a strong understanding of Julia's language semantics, and coding practices that differ substantially from those commonly adopted in the MATLAB/Simulink ecosystem. Nevertheless, the software-centric approach encouraged by Julia led to a modular architecture that is closer to flight software than traditional block-diagram-based simulators. Julia is a credible alternative for high-performance GNC simulation and prototyping, even if its possible adoption in an industrial context and beyond Phase 0/A/$\text{B}_1$ remains to be proven.

\section*{Acknowledgments}
The Authors would like to thank the entire ESA Argonaut Team, and more particularly Pedro Simplicio and Olivier Dubois-Matra for their support and contributions to ATLAS.

\end{document}